\documentstyle[aps,preprint]{revtex}
\begin{document}
\tightenlines

\title{Survival-extinction phase transition in a bit-string population with
mutation}

\author{Kathia M. Fehsenfeld,$^{1,*}$ Ronald Dickman,$^{1,\dagger}$ and
Am\'erico T. Bernardes$^{2,\ddagger}$}
\address{
$^1$Departamento de F\'{\i}sica, ICEx,
Caixa Postal 702,
Universidade Federal de Minas Gerais,
30123-970
Belo Horizonte - Minas Gerais, Brazil\\
$^2$Departamento de F\'{\i}sica,
Universidade Federal de Ouro Preto,
35400-000
Ouro Preto - Minas Gerais, Brazil
}

\date{\today}

\maketitle
\begin{abstract}
A bit-string model for the evolution of a population of haploid
organisms,
subject to competition,
reproduction with mutation and selection is studied, using mean field
theory and Monte Carlo simulations. We show that, depending on
environmental flexibility and genetic variability, the model
exhibits a phase transtion between extinction and survival.
The mean-field theory describes the infinite-size limit, while
simulations are used to study quasi-stationary properties.
\vspace{2em}

\noindent$^*$ Email address:  kathia@escelsa.com.br\\
$^\dagger$ Email address: dickman@fisica.ufmg.br \\
$^\ddagger$ Email address: atb@iceb.ufop.br\\

\end{abstract}
\vspace{1cm}

\newpage
\section{Introduction}

Many mathematical models have been proposed to describe the evolution
of populations, focusing on varied aspects, for example, mutation
accumulation \cite{Penna,Bernardes}, and adaptation
\cite{burger,charles,thoms,cebrat,pekalski}.
In the first case, how deleterious mutations
are passed to offspring, and the consequences for
individual growth, are of particular interest. In the
second, the principal interest is the
influence of different environmental conditions on the population.
One goal in this area is the development of a simple model capable of
describing the response of a population to environmental mutability.
Of interest, for example, is the ability of a population to adapt to
rapid changes in its environment.  Penna's bit-string model \cite{Penna}
seems well suited to this purpose.

In this paper, we propose a model for evolution of an adapting 
population,
to study the consequences of variation of conditions affecting survival,
related to environmental flexibility, and the genetic variability of
the population. Our main interest is to
describe the conditions determining the extinction or survival of
the population. 
The population evolves in discrete time with non-overlapping generations.
It consists of haploid organisms defined by their genotype (a
bit-string of $G$ positions, or genes). The individuals undergo asexual reproduction,
subject to mutation, competition and selection.
Selection is represented though a survival probability that depends on the
difference between a genome and a certain {\it ideal} genome.  Environmental
changes can be represented via alteration of this ideal.  In the present study,
however, the ideal genome is fixed, allowing a systematic study of the effect
of various other parameters upon survival.

We develop a mean-field (MFT) description, which 
describes the evolution of an infinite population exactly, since it has no spatial
structure. We also perform
Monte Carlo simulations for the model. The latter 
are useful for studying fluctuations due to finite population size,
that are not captured in the MFT.  We determine the survival/extinction phase boundary,
and compare the temporal evolution, and the genomic distribution of the population
predicted by MFT against simulation results.

The paper is organized as follows. In Sec. II, we define the model and in
Sec. III develop the MFT.
Sec. IV describes the Monte Carlo simulation
algorithm, while Sec. V reports MFT and simulation results.
We present our conclusions in Sec. VI.

\section{Model}

We study a model for evolution of a population of haploid individuals
defined by their genomes, and subject to competition, asexual reproduction with
mutation, and selection.
In this model, successive generations do not overlap. Each
individual is represented by a bit-string of $G$ positions (genes),
denoted by the vector
${\bf\sigma}=(\sigma_1,\sigma_2,...,\sigma_G)$, where
$\sigma_i=0$ or $1$.
The fitness of an individual to the environment
is measured in relation to a ``model individual" (or ``ideal genome"),
represented by the sequence $\sigma_i=0$, $i = 1,...,G$.
Each gene in state 1 represents a reduction in fitness, and carries the
same weight, independent of its position $i$.
Thus the Hamming distance from the ideal genome, given by $H = \sum_i
\sigma_i$,
characterizes an individual's fitness
(This manner of characterizing fitness
has been used in several studies of age-structured populations 
\cite{thoms,cebrat,pekalski}.)
The dependence of fitness on $H$ is through the {\it survival
probability}
\begin{equation}
S(H) = \frac{1+e^B}{e^{H/G \tau}+e^B}.
\label{survpr}
\end{equation}
$S(H)$ is the probability for an individual to survive up to the stage
in which she
must compete with the rest of the population; individuals that survive
the competition stage go on to reproduce offspring, as detailed below.
The parameter $\tau$, which plays a
role analogous to temperature in equilibrium statistical mechanics,
represents environmental flexibility, while $B$, which
is related to the genetic variability of the population, represents
mutational tolerance. $S(H)=1$ for $H=0$,
and decays monotonically with $H$.   We note that for fixed $H$ and $B$,
the survival probability is an increasing function of $\tau$, and that 
for
fixed $H$ and $\tau$, $S$ is an increasing function of $B$.   The 
Fermi-like
function $S(H)$ was used in a similar manner in the model of Thoms {\it 
et
al}
\cite{thoms}.  These authors define a death probability
$p_d  = [e^{\beta (b-a)}+1]^{-1}$,
where $\beta$ is an inverse temperature and $(b-a)$ represents
the difference between the typical number of mutations in the 
population,
and the number of mutations of the individual.

At reproduction, each organism is replaced by two offspring. The latter
are copies of their parent, with a certain number $m$ of mutations.
Each position has a probability of $\lambda$ to mutate
(mutations $0 \to 1$ and $1 \to 0$ are considered equally likely),
with mutations at different positions constituting independent events.
The number of mutations $m$ therefore follows a binomial distribution.
The mean number of mutations per reproduction event,
$\lambda G$, is set to unity in this study.

Competition amongst individuals is represented by the familiar
Verhulst factor,
\begin{equation}
V=1-\frac{N(t)}{N_{max}},
\label{hd}
\end{equation}
where $N(t)$ is the population at time $t$ and $N_{max}$ is the maximum
capacity of the environment.
The evolution of the population proceeds by discrete time steps:
at each step, the Verhulst factor
is applied by selecting at random (independently of $H$),
$NV$ survivors; the survivors go on to reproduce as described above.

\section{Mean-field theory}

We have developed a mean-field description
of the model defined above.  For this model, which has no spatial
structure, the deterministic mean-field description describes
the infinite-size limit ($N_{max} \to \infty$) exactly.
Differences between theory and
simulation are due to fluctuations
that appear in finite sized systems, but that are absent in the
infinite-size limit.

In the full stochastic description there are $2^G$ distinct genomes
$\sigma$, and an integer-valued random variable $N_\sigma(t) \geq 0$
for each.  Our first step in constructing a simplified
description is to reduce the set of variables to $N(H,t)$:
the number of individuals with Hamming distance $H$ from the ideal,
at time $t$.  Since the model
does not distinguish between individuals with the same Hamming
distance, the
probability distribution at any time $t>0$ will be a function of $H$
only,
if it is so at $t\!=\!0$.  We shall always suppose this to be the
case.

In the mean-field theory, the discrete-time evolution of the population
may be written so:
\begin{equation}
N(H,t\!+\!1) = {\sf E}[N(H,t\!+\!1)|\{N(H,t)\}],
\label{mft1}
\end{equation}
where $\{N(H,t)\}$ represents the entire set of population
variables at step $t$. In other words, the population at step
$t\!+\!1$ is approximated by its
{\it expected value}, given the distribution at step $t$.
(The latter, in turn, is given by the expected distribution, given
that for time $t\!-\!1$, and so on.)
The integer-valued random variables of the exact description
are therefore replaced by a set of real-valued, deterministic
variables.

Each step of the evolution consists of two stages: (1) death of 
individuals
due to competition for resources (`Verhulst stage');
(2) reproduction/selection.
In the Verhulst stage, the total population size $N \!=\! \sum_H N(H)$
is evaluated; then each subpopulation is reduced by the same factor,
$V = 1 \!-\! N/N_{max}$, yielding the values:
\begin{equation}
N'(H) = VN(H) , \;\;\; (H=0,...,G).
\label{verh}
\end{equation}
Note that the Verhulst stage involves an interaction between
individuals ($N'(H)$ is a nonlinear function of all of the $N(H)$),
and that each individual interacts equally with all others in this
process.

In the reproduction stage each individual is replaced by a pair of
offspring that have, in general, Hamming distances different from
those of the parent.  We assume independent, equally probable mutations
at each site, so that the number of mutations $m$ in a
given reproduction event is binomially distributed:
\begin{equation}
P(m) =
\left( \begin{array} {c}
 G \\
 m
\end{array} \right)
\lambda^m (1-\lambda)^{G-m} \;.
\label{binom}
\end{equation}
(Since $G \gg 1$ while the mean number of mutations $\lambda G$
is of order unity, we may approximate $P(m)$ by a Poisson distribution
in simulations; we retain the binomial distribution in the MFT analysis.)

Each reproduction event may be represented schematically as $H' \to H_1,
H_2$,
where $H'$ denotes the Hamming distance of the parent and $H_1$ and 
$H_2$
those of
the offspring.  Since $H' \to H_1$ and $H' \to H_2$ are independent 
events
(even though they happen simultaneously), it suffices to consider
one such, i.e., $H' \to H$; let $W(H|H')$ represent its probability.
If the offspring differs from its parent at exactly $m$ positions, then,
\[
\max[0,H'-m] \leq H \leq \min[H'+m,G].
\]
Let $m = m_0 + m_1$, with $m_0$ the number of mutations $0 \to 1$ and
$m_1$ the number of type $1 \to 0$.   Each event
is characterized by $H'$, $m$, and $m_0$.
(Evidently, $H = H' + m_0 - m_1 = H' + 2m_0 - m$.)
The probability of such an event is given by the
hypergeometric distribution:
\begin{equation}
p(m_0|m,G,H') = \frac{
\left( \begin{array} {c}
 G-H' \\
 m_0
\end{array} \right)
\left( \begin{array} {c}
 H' \\
 m-m_0
\end{array} \right) }
{\left( \begin{array} {c}
 G \\
 m
\end{array} \right) }
\label{hiper} \;.
\end{equation}
Now using $m_0 = (H\!-\!H'\!+\!m)/2$, we have,
\begin{equation}
W(H|H') = (G\!-\!H')! H'! \sum_{m=0}^G \frac{ \lambda^m 
(1-\lambda)^{G-m} }
{\left(\frac{H-H'+m}{2} \right)!
\left(\frac{H'-H+m}{2} \right)!
\left(G\!-\!\frac{H+H'+m}{2} \right)!
\left(\frac{H'+H-m}{2} \right)!} \;.
\label{w2}
\end{equation}
Next we observe that the expected number of {\it surviving} offspring
with Hamming distance $H$ produced by a parent with Hamming
distance $H'$ is: $\tilde{W}(H|H') \equiv 2S(H)W(H|H')$.
Thus the expected number of individuals with Hamming
distance $H$, at step $t\!+\!1$ is:
\begin{equation}
{\sf E} [N(H,t\!+\!1)|\{N(H',t)\}] =
\sum_{H'=0}^G \tilde{W}(H|H') N'(H') \;,
\label{eqtcm}
\end{equation}
where $N'(H')$ is the distribution just after the Verhulst step.
The evolution of the population is found via numerical iteration of Eqs.
(\ref{verh}) and (\ref{eqtcm}).

\section{Simulation algorithm}

We study the evolution of the model population in Monte Carlo simulations.
Initially, $N_0 = N_{max}/10$ individuals of $G=128$ bits are 
generated,
each with a random gene sequence,
${\bf\sigma}=(\sigma_1,\sigma_2,...,\sigma_G)$, where
$\sigma_i=0$ or $1$ with equal likelihood.
The procedure is as follows:

{\it i)} The Verhulst factor $V = 1 - N(t)/N_{max}$ is evaluated.
Then for each individual,
a random number $s$ is generated; the individual survives
(dies) if $s<V$ ($s>V$).

{\it ii)} Each individual reproduces: 2 copies are created, with
possible
mutations.  The number of mutations $m$ is given by a random integer,
chosen from a Poisson distribution with parameter 1.
The mutation loci are selected at random.

{\it iii)} For each daughter, the Hamming distance $H$ from the ideal
is evaluated, and
a random number $r$, uniform on [0,1] is generated.
If $r \leq S(H)$, the individual survives; otherwise, it dies.

During the simulations, we record the population, average Hamming
distance, the average survival probability,
\begin{equation}
\langle S(t) \rangle = \frac{1}{N(t)} \sum_{i=1}^{N(t)} S(H_i),
\end{equation}
and the {\it survival rate}, ${\cal S}(t) \equiv N(t)/N(t-1)$.
(Note that in general $\langle S(t) \rangle < 1$, while ${\cal S}(t)$
may, in principle, take any nonnegative value, and is unity in the
stationary state.)
Depending on the
parameters $\tau$, $B$, and $N_{max}$, the population may survive until
a certain maximum time
($t_{max} = 30\;000$ steps in the simulations), attaining a quasi-stationary state, or
may go
extinct.
We record the Hamming distance distribution in the
quasi-stationary state.

\section{Results and Discussion}

Depending on the values of $B$ and $\tau$
that characterize the survival probability function $S(H)$,
Eq. (\ref{survpr}), the
population either survives or goes extinct.  In the mean-field
theory this is a sharp transition. In simulations, due to
finite population size, fluctuations into the absorbing state
(population zero) are to be expected.  Indeed, for any {\it finite}
system size the population must eventually go extinct, if the
process is permitted to continue indefinitely.  We adopt
$t_{max} = 30\;000$ as a convenient maximum time, allowing us to
discriminate between survival and extinction, and (in the former case),
study quasi-stationary properties, except very near the
transition, where, as noted, the sharp distinction is blurred by
fluctuations.

Fig. 1 shows the phase boundary between survival and extinction in the
$B$ - $\tau$ plane, comparing the mean-field prediction against
simulations using $N_{max}$ = $10^4, 10^5$ and $5 \times 10^5$.
As $N_{max}$ is increased, the survival/extinction line found in
simulation approaches the MFT prediction, as expected.
For small values of $\tau$, (a ``hard" or inflexible
environment), survival of population requires high values of $B$, the
mutational tolerance.
The mean-field survival/extinction line of the diagram is obtained
by fixing the parameter $\tau$ and measuring the stationary
population density $\rho = N/N_{max}$ as a function of $B$.
Near the transition, $\rho$ depends linearly on $B$:
$\rho \propto B \!-\! B_c(\tau)$, as is normally the case
in mean-field descriptions of a continuous phase transition
to an absorbing state \cite{marro}.
The line $B_c(\tau)$ is readily
obtained via linear regression to the $\rho(B)$ data near the
transition.
Note that $B_c = 0$ for
$\tau > 0.192 $.  For $\tau \ll 1$, on the other hand,
$B_c \propto 1/\tau$.  (Increasing the mutation probability
$\lambda$, the phase boundary is displaced upward and to
the right, enlarging the extinction region.)
Fig. 2 is a three-dimensional plot of the population density
as a function of $B$ and $\tau$;
the extinction region is evident, as is the monotonic
growth of $\rho$ with either parameter.

Fig. 3 presents a typical evolution of the population density
$\rho(t)$.
For $B$ and $\tau$ in the
survival phase, the population exhibits a rapid initial decay
and then evolves to a quasi-stationary state.
Simulation and MFT evolutions are in good agreement,
despite fluctuations in the former.

The quasi-stationary distribution of Hamming distances obtained
in simulation is compared in Fig. 4 with the stationary distribution
predicted by mean-field theory. In all
cases, the distribution peaks near the mean value $<H>$, and
has a generally Gaussian appearance. For fixed $\tau$, we observe that
$<H>$ increases monotonically with $B$, attaining a {\it plateau}, if 
$\tau$ is sufficiently large.
The plateau value is $<H> \simeq 64$, i.e.,
half the genome size. For fixed $B$,
we observe that $<H>$ increases with $\tau$,
until attaining $<H>=64$.
The variance of the distribution behaves similarly.
Its saturation value is about 32, giving
a standard deviation $\sigma \simeq 5.7$.
This is not surprising, given that $B$ and $\tau$
both represent tolerance of differences from the ideal genome.
Fig. 5 shows the stationary values of $<H>$ and $\sigma_H$ as
functions of $B$, as predicted
by MFT; simulations yield very similar behaviour.
In simulations, extinction occurs at larger $B$ values
than are predicted by MFT, due to finite-size
effects, as noted above; the difference between simulation and
theory diminishes with increasing system size.

\section{Summary}

We propose a bit-string model of the evolution a simple haploid 
population.
Similarly to previous studies \cite{thoms,cebrat,pekalski}, the model
includes the effect of enviromental flexibility and tolerance to genetic
differences on the survival probability.
Unlike previous works, we employ a survival probability
that is a monotonic increasing function of the
parameters $B$ and $\tau$ that represent tolerance of
genetic difference between a given genome and
the ideal. The model is studied via
computer simulations and mean-field theory, which are
in good agreement.

The model, like many others in
population dynamics or epidemic analysis,
exhibits a continuous transition between an active phase (survival)
and an absorbing one (extinction).  We map out the phase boundary in
the $B$ - $\tau$ plane, and find clear evidence of mean-field-like
critical behavior, as in other population models lacking spatial
structure \cite{marro}.  The mean-field description is
exact in the infinite-size limit, but provides no information
regarding fluctuations.  On the other hand,
simulations for parameter values in the
active phase yield information on the quasi-stationary state of a
finite system ($N_{max} < \infty$).  It is also of interest to obtain
the {\it lifetime} of this quasi-stationary state, or, equivalently,
the mean first-passage time to extinction.  Such information can
in principle be obtained from simulations, or from a probabilistic
analysis of finite populations, starting from the master equation
\cite{qss}.  Given the large number of random variables involved
($G\!+\!1$, if we assume that the probability depends only on Hamming
distance $H$), the multivariate Fokker-Planck equation would seem
the most convenient tool; theoretical analysis of finite populations
is left as subject for future work.  The simulation
results reported here should prove useful in testing such theories.

Another interesting direction for future study is
the response of the population to changes in the environment.
Such changes can be represented by variations in the
ideal genome (as presented in \cite{cebrat,pekalski}) 
and/or in the parameters $\tau$, $B$, $\lambda$,
and $N_{max}$.
A related question is that of transitions
in the genome distribution when two or more ideals (corresponding
to distinct, well adapted types in the fitness landscape), exist.
Studies of these problems using the bit-string model are in progress.
\vspace{1em}

\noindent{\bf Acknowledgments}

\noindent ATB acknowledges the kind hospitality of the Departamento de
F\'{\i}sica-UFMG. This work was
partially supported by the Brazilian Agencies CNPq, FINEP and FAPEMIG.

\newpage
\noindent{\bf Figure Captions}
\vspace{1em}

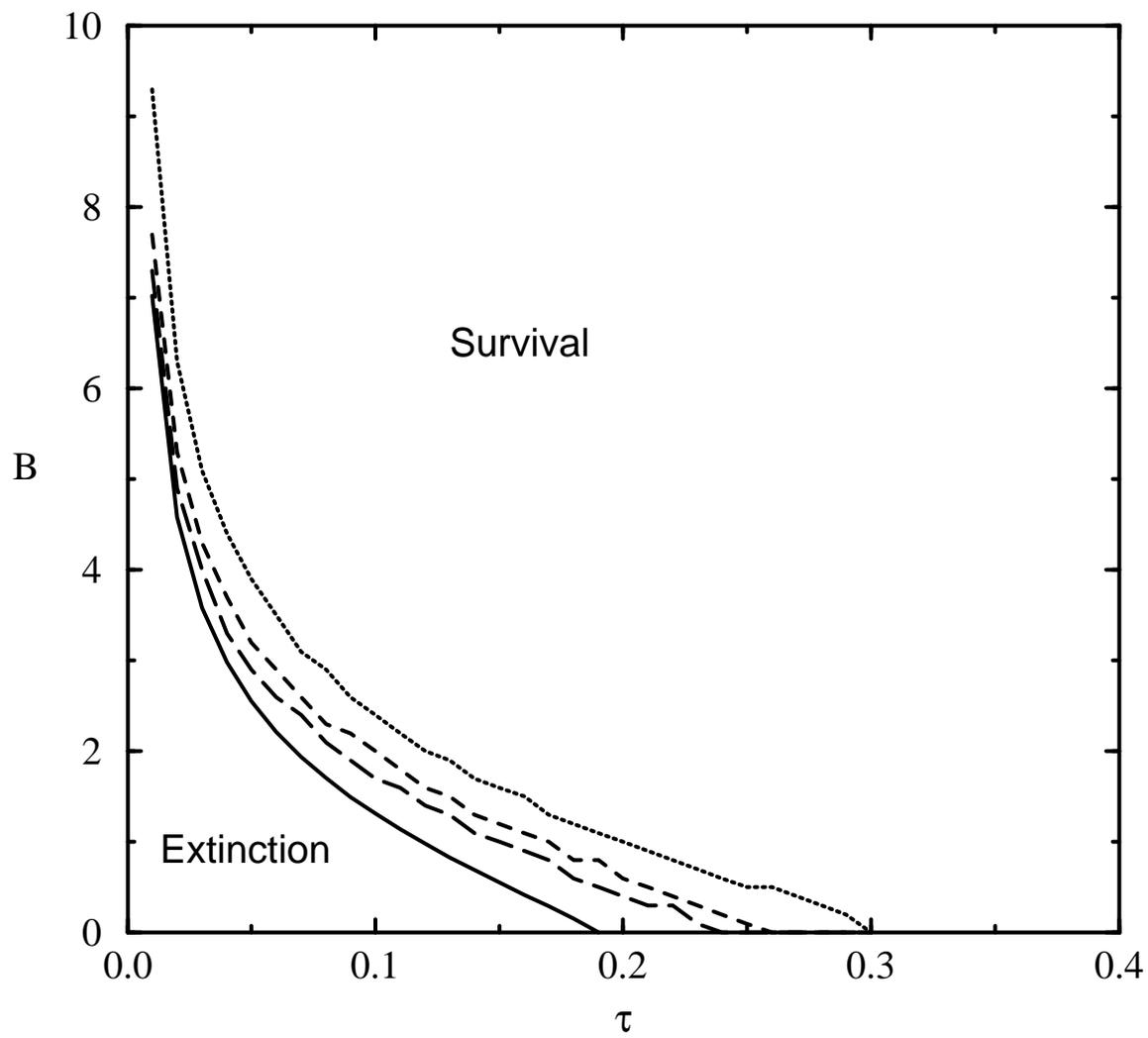
\noindent Fig. 1.
Survival/extinction phase boundary in the $B$-$\tau$ plane for $\lambda
G = 1$.
The solid line is the MFT prediction;
dashed lines represent simulation results for
$N_{max}=5 \times 10^5$, $10^5$ and
$10^4$ (bottom to top).
\vspace{1em}

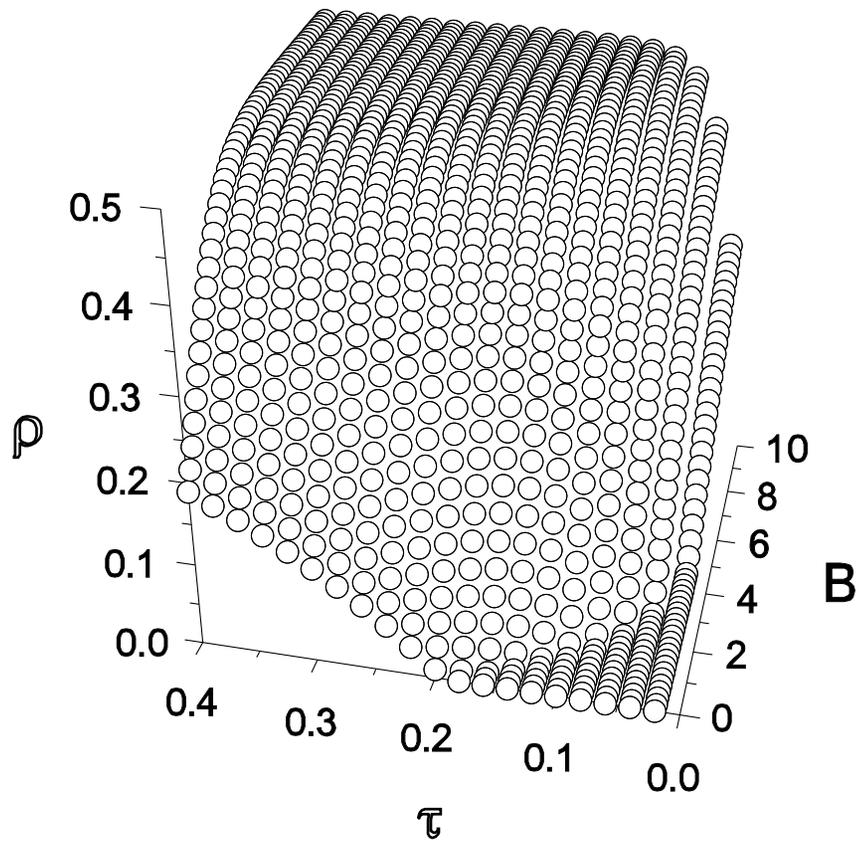
\noindent Fig. 2.
Population density
$\rho$ as a function of $B$ and $\tau$ from MFT.
For $\tau \geq 0.192$, the population survives for
any value of $B$.
\vspace{1em}

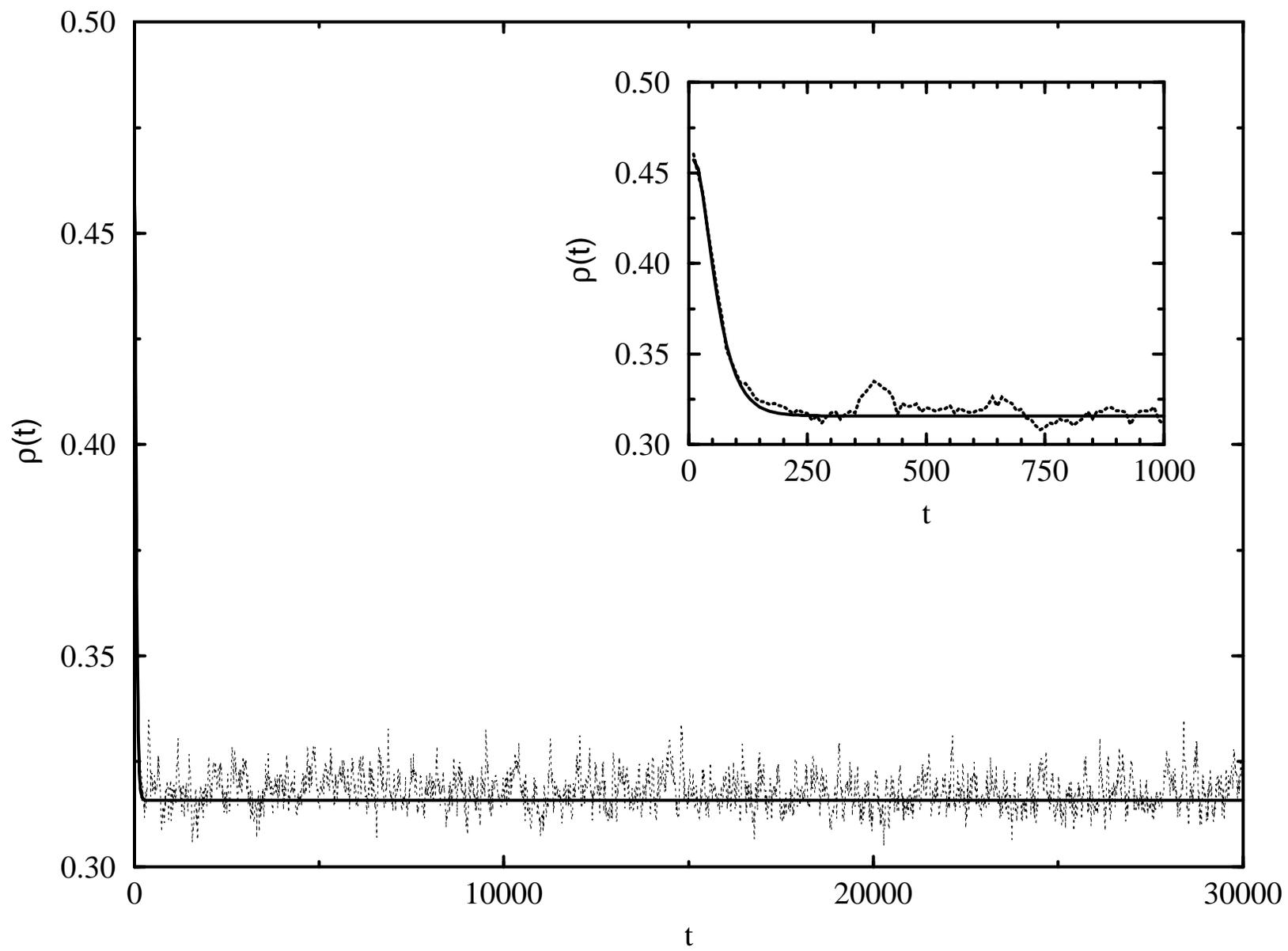
\noindent Fig. 3.
Time evolution of the population density $\rho$
for $\tau=0.1$ and $B=4$,
in MFT (smooth curve) and simulation ($N_{max} = 10^5$).
\vspace{1em}

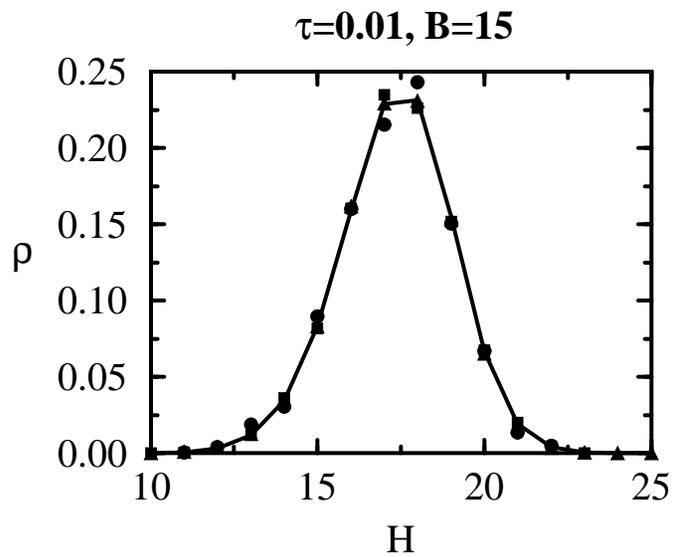
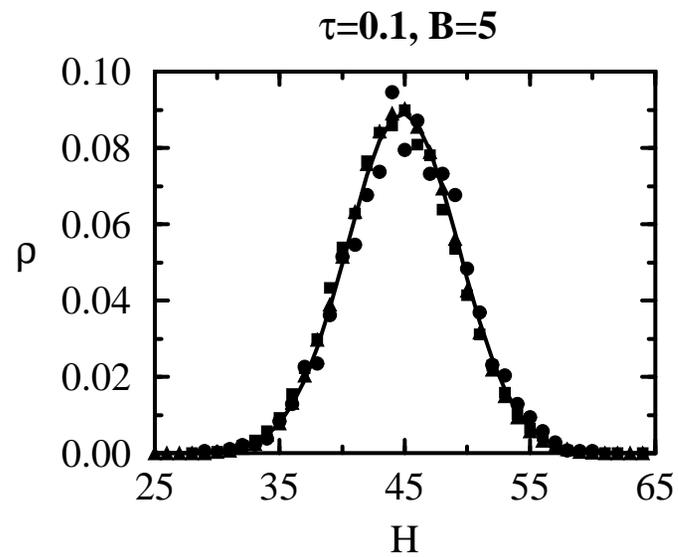
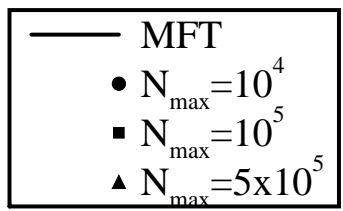
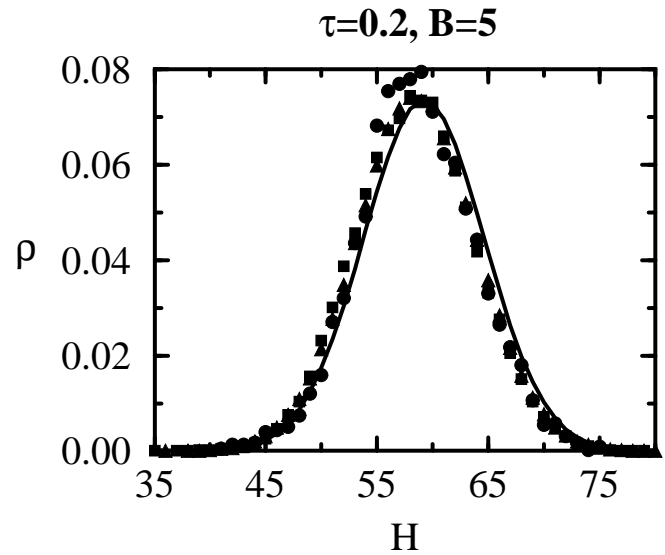
\noindent Fig. 4.
Stationary
Hamming-distance distribution
for various parameters, as indicated.
\vspace{1em}

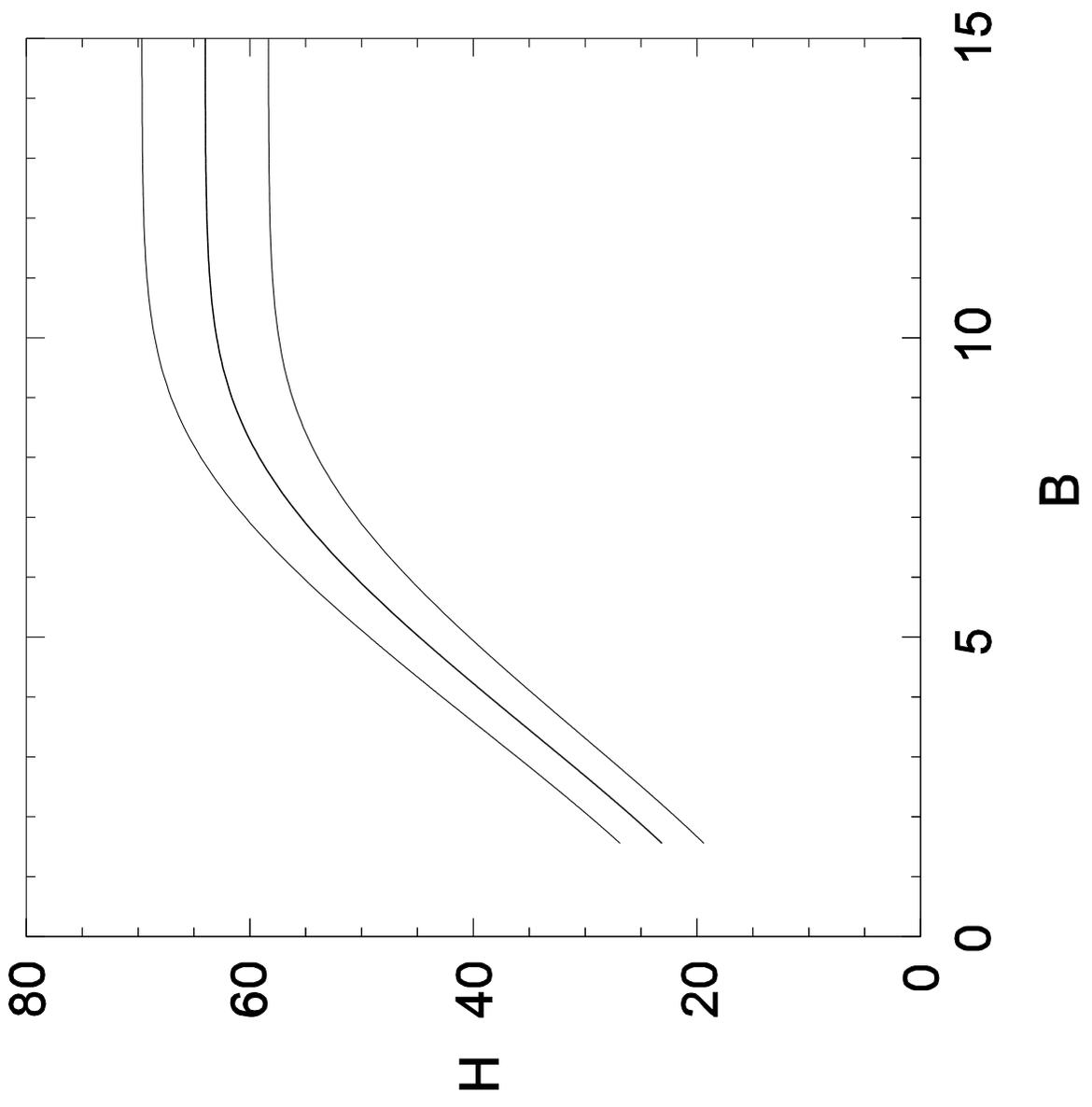
\noindent Fig. 5.
Dependence of Hamming distance on $B$ for $\tau = 0.1$ in MFT. Central line:
mean Hamming distance, $<H>$; upper and lower lines
represent one standard deviation above or below the mean.
\vspace{1em}

\end{document}